\documentclass[twocolumn,amsmath,amssymb]{revtex4}
\usepackage{amsmath,amssymb,graphicx}
\usepackage{graphicx}
\usepackage{dcolumn}
\usepackage{bm}
\usepackage{epsfig}
\usepackage{here}
\usepackage{float}
\usepackage{amsmath}
\usepackage[usenames]{color}
\usepackage{amsmath}
\usepackage{epsfig}
\usepackage{color}
\usepackage{soul}

\newcommand{\bea}{\begin{eqnarray}}
\newcommand{\eea}{\end{eqnarray}}

\begin{document}
%%%%%%%%%%%%%%%%%%%%%%%%%%%%%%%%%%%%%%%%%%%%%%%%%%%%%%%%%%%%%%%
\title{Pulsar Timing Array Signature from Oscillating Metric Perturbations due to Ultra-light Axion}
\author{Jai-chan Hwang${}^{1}$, Donghui Jeong${}^{2, 3}$, Hyerim Noh${}^{4}$, Clemente Smarra${}^{5, 6}$}
\address{${}^{1}$Particle Theory and Cosmology Group,
         Center for Theoretical Physics of the Universe,
         Institute for Basic Science (IBS), Daejeon, 34126, Republic of Korea
         \\
         ${}^{2}$Department of Astronomy and Astrophysics and Institute for Gravitation and the Cosmos,
         The Pennsylvania State University, University Park, PA 16802, USA
         \\
         ${}^{3}$School of Physics, Korea Institute for Advanced Study (KIAS), 85 Hoegiro, Dongdaemun-gu, Seoul, 02455, Republic of Korea
         \\
         ${}^{4}$Theoretical Astrophysics Group, Korea Astronomy and Space Science Institute, Daejeon, Republic of Korea
         \\
         ${}^{5}$SISSA --- International School for Advanced Studies, Via Bonomea 265, 34136, Trieste, Italy and INFN, Sezione di Trieste
         \\
         ${}^{6}$IFPU --- Institute for Fundamental Physics of the Universe, Via Beirut 2, 34014 Trieste, Italy
         }

%%%%%%%%%%%%%%%%%%%%%%%%%%%%%%%%%%%%%%%%%%%%%%%%%%%%%%%%%%%%%%%
\date{\today}

%%%%%%%%%%%%%%%%%%%%%%%%%%%%%%%%%%%%%%%%%%%%%%%%%%%%%%%%%%%%%%%
\begin{abstract}

A coherently oscillating ultra-light axion can behave as dark matter. In particular, its coherently oscillating pressure perturbations can source an oscillating scalar metric perturbation, with a characteristic oscillation frequency which is twice the axion Compton frequency. A candidate in the mass range $10^{(-24,-21)}{\rm eV}$ can provide a signal in the frequency range tested by current and future Pulsar Timing Array (PTA) programs. Involving the pressure perturbations in a highly nonlinear environment, such an analysis demands a relativistic and nonlinear treatment. Here, we provide a rigorous derivation of the effect assuming weak gravity and slow-motion limit of Einstein's gravity in zero-shear gauge and show that dark matter's velocity potential determines the oscillation phase and frequency change. A monochromatic PTA signal correlated with the velocity field would confirm the prediction, for example, by cross-correlating the PTA results with the future local velocity flow measurements.

\end{abstract}

%%%%%%%%%%%%%%%%%%%%%%%%%%%%%%%%%%%%%%%%%%%%%%%%%%%%%%%%%%%%%%%
\maketitle
%\tableofcontents

%%%%%%%%%%%%%%%%%%%%%%%%%%%%%%%%%%%%%%%%%%%%%%%%%%%%%%%%%%%%%%%
\section{Introduction}

A novel way of observationally probing the ultra-light axion as a fuzzy dark matter candidate was proposed by Khmelnitsky and Rubakov \cite{Khmelnitsky-Rubakov-2014}. The coherent oscillation of ultra-light axion dark matter leads to pressure perturbations oscillating at twice its Compton frequency and consequently to oscillation in gravitational potential. It was suggested that this effect can produce distinctive features in the travel time of radio beams emitted from pulsars, which have been monitored for decades in Pulsar Timing Array (PTA) experiments. The derivation in \cite{Khmelnitsky-Rubakov-2014}, however, is incomplete by not determining the phase of the axion field oscillation, which can indeed be addressed using the equation of motion of the axion.

Previously we analysed the case using linear perturbation theory in cosmological context \cite{Hwang-Noh-2022-oscillation}. Linear perturbation study is not suitable for the purpose considered in \cite{Khmelnitsky-Rubakov-2014} as pulsars are located within galactic halo or disk surrounded by the dark matter where density enhancements are huge, i.e., nonlinear in density distribution, compared with the background cosmological medium. Meanwhile inhomogeneity in gravitational potential is known to be extremely small in all cosmological scales \cite{Kim-2022-CP}.

Here, we derive the gravitational potential oscillation and density perturbation equations in an environment of highly non-linear dark matter density enhancement, in non-expanding background. Oscillation of the gravitational potential caused by the pressure perturbation is a pure relativistic effect operating in Newtonian order, see later. We consider the weak-gravity and slow-motion approximation, but consider nonlinear matter and field inhomogeneities.

We {\it develop} the weak-gravity formulations for the fluid and field in Secs.\ \ref{sec:WG-SM} and \ref{sec:MSF}, respectively, and apply the formulations to the coherently oscillating ultra-light axion in Sec.\ \ref{sec:axion}. The phase is determined by the perturbed velocity potential of the axion fluid which include the position dependent phase and the post-Newtonian order frequency shift. Implications on potential PTA measurement is made in Sec.\ \ref{sec:PTA}. We discuss our result in Sec.\ \ref{sec:discussion}.

%%%%%%%%%%%%%%%%%%%%%%%%%%%%%%%%%%%%%%%%%%%%%%%%%%%%%%%%%%%%%%%
\section{Weak-gravity and slow-motion approximation: fluid}
                                   \label{sec:WG-SM}

We consider the relativistic fully nonlinear and exact perturbation equations in the cosmological context \cite{Hwang-Noh-2013, Hwang-Noh-Park-2016}. On a flat Friedmann background, the metric is
\bea
   & & g_{00}
       = - a^2 \left( 1 + 2 \alpha \right), \quad
       g_{0i} = - a \chi_{i},
   \nonumber \\
   & &
       g_{ij} = a^2 \left( 1 + 2 \varphi \right) \delta_{ij},
   \label{metric}
\eea
where the scale factor $a(t)$ is a function of time only but $\alpha$, $\varphi$ and $\chi_i$ are functions of space and time with arbitrary amplitudes; the spatial indices are raised and lowered using $\delta_{ij}$ as the metric; in Minkowski background, we set $a \equiv 1$. This metric indicates that we have only five degrees of freedom; we {\it ignored} the transverse-tracefree tensor-type perturbation (two physical degrees of freedom), and {\it imposed} a spatial gauge condition (three gauge degrees of freedom, without losing generality to fully nonlinear order) \cite{Hwang-Noh-2013}. The temporal gauge (slicing or hypersurface) condition is not imposed yet.

We further {\it ignore} transverse vector-type perturbation (two physical degrees of freedom), thus $\chi_i = \chi_{,i}$, and we {\it take} zero-shear gauge, setting the longitudinal part of $\chi_i$ equal to zero, i.e., $\chi \equiv 0$ as the slicing condition, thus $\chi_i = 0$. Under this temporal gauge condition, together with the already chosen spatial gauge condition, all remaining variables are free from the gauge degrees of freedom and are equivalently (spatially and temporally) gauge invariant, meaning that the gauge degrees of freedom are completely fixed with no remnant gauge mode to fully nonlinear order \cite{Bardeen-1988, Hwang-Noh-2013}.

The energy momentum tensor is decomposed into fluid quantities based on the time-like four-vector $u_a$ as \cite{Ellis-1971}
\bea
   T_{ab}
       = \mu u_a u_b
       + p \left( g_{ab} + u_a u_b \right)
       + q_a u_b + q_b u_a + \pi_{ab},
   \label{Tab-fluid}
\eea
with $q^a u_a \equiv 0 $, $\pi_{ab} u^b = 0 = \pi^a_a$, and $u^a u_a \equiv -1$. Thus, we have
\bea
   & & \mu \equiv T_{ab} u^a u^b, \quad
       p \equiv {1 \over 3} T_{ab} h^{ab}, \quad
       q_a \equiv - T_{cd} u^c h^d_a,
   \nonumber \\
   & &
       \pi_{ab} \equiv T_{cd} h^c_a h^d_b - p h_{ab},
   \label{fluid-quantities}
\eea
where $h_{ab} \equiv g_{ab} + u_a u_b$ is the spatial-projection tensor. Fluid quantities $\mu$, $p$, $q_a$, and $\pi_{ab}$ are the energy density, pressure, flux vector, and anisotropic stress, respectively, measured by an observer with the four-velocity $u_a$; we define $\mu = \varrho c^2$. The four-vector becomes
\bea
   u_i = {v_i \over c}, \quad
       u_0 = - 1 - \alpha, \quad
       u^i = {v^i \over c}, \quad
       u^0 = 1 - \alpha.
   \label{four-vector}
\eea
For a fully nonlinear form, see Eq.\ (29) in \cite{Hwang-Noh-Park-2016}.

Our main approximation in the following is the {\it weak-gravity} limit ($\alpha \ll 1$ and $\varphi \ll 1$) and {\it slow-motion} approximation ($v^i v_i/c^2 \ll 1$); in this way, we consider only {\it linear} order terms involving the gravity ($\alpha$ and $\varphi$).

Under the conditions mentioned, and in Minkowski background setting $a \equiv 1$, Eqs.\ (6)-(10) and (14)-(17) in \cite{Hwang-Noh-Park-2016} reduce to
\bea
   & & \kappa \equiv 3 {\dot \Psi \over c^2},
   \label{eq1-M} \\
   & & \Delta \Psi = 4 \pi G \left[ \delta \varrho
       + \left( \varrho
       + {p \over c^2} \right)
       \left( \gamma^2 - 1 \right) \right],
   \label{eq2-M} \\
   & & \kappa = - {12 \pi G \over c^2}
       \Delta^{-1} \nabla_i \left[
       \left( \varrho
       + {p \over c^2} \right) v^i \right],
   \label{eq3-M} \\
   & & \dot \kappa + \Delta \Phi
       - {12 \pi G \over c^2}
       \left( \bar \varrho + {\bar p \over c^2} \right) \Phi
       = 4 \pi G \left( \delta \varrho
       + 3 {\delta p \over c^2} \right),
   \label{eq4-M} \\
   & & \Psi = \Phi,
   \label{eq5-M} \\
   & & \delta \dot \varrho
       + v^i \nabla_i \left( \delta \varrho
       - {\delta p \over c^2} \right)
       + \left( \varrho + {p \over c^2}
       \right) \left( v^i_{\;,i} - \kappa \right)
       = 0,
   \label{eq6-M} \\
   & & \dot v_i + v^j \nabla_j v_i
       + \Phi_{,i}
       + {1 \over \varrho + p/c^2}
       \bigg( \delta p_{,i}
       + {\dot p \over c^2} v_i \bigg) = 0,
   \label{eq7-M}
\eea
where we used
\bea
   \alpha \equiv {\Phi \over c^2}, \quad
       \varphi \equiv - {\Psi \over c^2},
\eea
and set $\varrho ({\bf x}, t) \equiv \bar \varrho (t) + \delta \varrho ({\bf x}, t)$ with no restriction on the amplitude of $\delta \varrho$, and similarly for $p$; $\bar \varrho$ is the background density. In the fluid formulation, for simplicity, we {\it ignore} the anisotropic stress. Although anisotropic stress is not supported by the scalar field, it appears in the axion case to nonlinear order \cite{Noh-Hwang-Park-2017}; we will address this issue later, see below Eq.\ (\ref{EOM-MSF-E-frame}).

To the background order we have $\dot {\bar \varrho} = 0$. We note that even in Minkowski background, it is important to subtract the background order (Friedmann) equations properly; this applies to Eqs.\ (\ref{eq2-M}) and (\ref{eq4-M}). In this way, the potentials $\Phi$ and $\Psi$ are perturbed order, and the situation is consistent with Jeans' choice (often know as Jeans swindle) \cite{Jeans-1902}.

In deriving Eqs.\ (\ref{eq6-M}) and (\ref{eq7-M}) we should carefully consider the time derivative of the Lorentz factor. Using Eq.\ (\ref{eq7-M}), we have $( \mu + p ) \dot \gamma = - v^i \nabla_i \delta p$ to our approximation; for later use of this property we kept $(\gamma^2 - 1)$-term in Eq.\ (\ref{eq2-M}). Combining Eqs.\ (\ref{eq6-M}) and (\ref{eq7-M}), we have
\bea
   & & \left[ \left( \varrho
       + {p \over c^2} \right) v_i
       \right]^{\displaystyle{\cdot}}
       + \left[ \left( \varrho
       + {p \over c^2} \right) v_i v^j \right]_{,j}
   \nonumber \\
   & & \qquad
       + \nabla_i \delta p
       + \left( \bar \varrho + {\bar p \over c^2} \right)
       \nabla_i \Phi
       = 2 v_i v^j \nabla_j {\delta p \over c^2}.
   \label{eq7-M-2}
\eea
We can estimate $G \varrho/(c^2 \Delta) \sim (\ell/c)^2/t_g^2 \sim G M/(\ell c^2) \sim \Phi/c^2$ with $\Delta \sim 1/\ell^2$, $M \sim \varrho \ell^3$ and $t_g \sim 1/\sqrt{G \varrho}$; $\ell$ and $M$ are characteristic length and mass scale, respectively, and $t_g$ the gravitational time scale. Thus, Eq.\ (\ref{eq3-M}) shows that $\kappa$ is negligible compared with $\nabla \cdot {\bf v}$ under the weak-gravity condition or the action-at-a-distance condition, see \cite{Hwang-Noh-2016}.

Equations (\ref{eq6-M}), (\ref{eq7-M}) and (\ref{eq2-M}), using Eq.\ (\ref{eq5-M}), give
\bea
   & & \delta \dot \varrho
       + \nabla \cdot \left[
       \left( \varrho
       + {p \over c^2} \right) {\bf v} \right]
       = {2 \over c^2} {\bf v} \cdot \nabla \delta p,
   \label{E-conserv-M} \\
   & & \dot {\bf v} + {\bf v} \cdot \nabla {\bf v}
       + \nabla \Phi
       = - {1 \over \varrho + p/c^2}
       \bigg( \nabla \delta p
       + {\dot {p} \over c^2} {\bf v} \bigg),
   \label{Mom-conserv-M} \\
   & & \Delta \Phi = 4 \pi G \delta \varrho.
   \label{Poisson-eq-M}
\eea
With the equation of state provided, these are the closed set of equations describing a fluid in weak-gravity and slow-motion limit (i.e., non-relativistic velocity) but with relativistic pressure \cite{Hwang-Noh-2013-pressure}. In the non-relativistic pressure limit we recover the well known non-relativistic fluid equation with Newtonian gravity.

Notice the absence of pressure term in the Poisson equation which leads to {\it contradiction} with the exact case in the spherically symmetric medium \cite{Tolman-1939}. The pressure term in Poisson equation with $\delta \varrho \rightarrow \delta \varrho + 3 \delta p/c^2$ is recovered in the maximal slicing (the uniform-expansion gauge setting $\kappa \equiv 0$), but the term is missing in zero-shear gauge, see \cite{Hwang-Noh-2016}.

We still have Eqs.\ (\ref{eq1-M}), (\ref{eq3-M}) and (\ref{eq4-M}) remaining. Previously, in \cite{Hwang-Noh-2013-pressure, Hwang-Noh-2016} we have used these equations only to check the consistency of our weak-gravity approximation combined with relativistic matter. Using Eq.\ (\ref{eq3-M}) together with Eq.\ (\ref{eq7-M-2}), Eq.\ (\ref{eq4-M}) gives Eq.\ (\ref{Poisson-eq-M}). The final remaining one in Eq.\ (\ref{eq1-M}) can be checked as follows. Here, we have to take the time derivative of Eq.\ (\ref{eq2-M}). Remembering the contribution from $\dot \gamma$ mentioned above Eq.\ (\ref{eq7-M-2}), using Eqs.\ (\ref{eq2-M}) and (\ref{E-conserv-M}), we can show Eq.\ (\ref{eq1-M}) gives Eq.\ (\ref{eq3-M}). This {\it proves} the full consistency of the weak field slow-motion approximation equations with relativistic pressure.

On the other hand, using Eq.\ (\ref{eq1-M}) together with Eqs.\ (\ref{eq2-M}) and (\ref{eq5-M}), Eq.\ (\ref{eq4-M}) gives
\bea
   \ddot \Psi = 4 \pi G \left[ \delta p
       + \left( \bar \mu + \bar p \right) \alpha \right].
   \label{ddot-Psi-M}
\eea
This is another relation, noticed in \cite{Khmelnitsky-Rubakov-2014}, and the main focus of this work. Therefore, considering the relativistic pressure, Eqs.\ (\ref{E-conserv-M})-(\ref{ddot-Psi-M}) provide a complete set.

These equations are derived by reducing the fully nonlinear and exact perturbation equations in Einstein's gravity \cite{Hwang-Noh-2013, Hwang-Noh-Park-2016}, and we have {\it not} imposed conditions on amplitudes of $\delta \mu$ and $\delta p$; this implies that we can consider {\it nonlinear} density and pressure perturbation as the source of the gravity. We note that this is {\it different} from the post-Newtonian expansion in \cite{Chandrasekhar-1965}. The weak-gravity approximation was developed to include fully relativistic, thus without assuming slow-motion, hydrodynamic matter in \cite{Hwang-Noh-2016}.

To the {\it linear} order, from Eqs.\ (\ref{E-conserv-M})-(\ref{Poisson-eq-M}) we can derive
\bea
   \ddot \delta - 4 \pi G \bar \varrho \delta
       = {\Delta \delta p / \bar \varrho},
   \label{ddot-delta-eq}
\eea
where $\delta \equiv \delta \varrho/\bar \varrho$; we used $\dot {\bar \varrho} = 0$ but we have {\it not} imposed any condition on the pressure.

The above fluid formulation can be used when a scalar field constitutes the fluid. What we need are fluid quantities, like $\delta \mu$, $\delta p$ and ${\bf v}$, expressed in terms of the field, and we additionally have the equation of motion which can replace (or complement) the conservation equations. These are derived in the next section.

%%%%%%%%%%%%%%%%%%%%%%%%%%%%%%%%%%%%%%%%%%%%%%%%%%%%%%%%%%%%%%%
\section{Scalar field}
                                   \label{sec:MSF}

Here, we {\it derive} fluid quantities and the equation of motion for a minimally coupled scalar field with a general potential first in the covariant form, and then in the weak-gravity and slow-motion limit. As a result, when the scalar field constitutes the fluid, Eqs.\ (\ref{fluid-MSF-WG}) and (\ref{EOM-MSF-WG}) supplement the fluid formulation in the same limit presented in Sec.\ \ref{sec:WG-SM}.

%%%%%%%%%%%%%%%%%%%%%%%%%%%%%%%%%%%%%%%%%%%%%%%%%%%%%%%%%%%%%%%
\subsection{Covariant forms}
                                   \label{sec:MSF-covariant}

The energy-momentum tensor and equation of motion for the scalar field are
\bea
   & & T_{ab}
       = \phi_{,a} \phi_{,b}
       - \left( {1 \over 2} \phi^{;c} \phi_{,c} + V
       \right) g_{ab},
   \label{Tab-MSF} \\
   & & \Box \phi = V_{,\phi}.
   \label{EOM-MSF}
\eea
Using Eq.\ (\ref{Tab-MSF}), the fluid quantities in Eq.\ (\ref{fluid-quantities}) become
\bea
   & & \mu = {1 \over 2} \widetilde {\dot \phi}{}^2
       + V + {1 \over 2} h^{ab} \phi_{,a} \phi_{,b},
   \nonumber \\
   & &
       p = {1 \over 2} \widetilde {\dot \phi}{}^2
       - V - {1 \over 6} h^{ab} \phi_{,a} \phi_{,b}, \quad
       q_a = - \widetilde {\dot \phi} h^b_a \phi_{,b},
   \nonumber \\
   & &
       \pi_{ab} = h_a^c \phi_{,c} h_b^d \phi_{,d}
       - {1 \over 3} h_{ab} h^{cd} \phi_{,c} \phi_{,d},
   \label{fluid-MSF-cov}
\eea
where $\widetilde {\dot \phi} \equiv \phi_{,c} u^c$.
The equation of motion in Eq.\ (\ref{EOM-MSF}) gives
\bea
   \widetilde {\ddot \phi}
       + \theta \widetilde {\dot \phi}
       + V_{,\phi}
       - h_a^b \left( h^{ac} \phi_{,c} \right)_{;b}
       - h_a^b \phi_{,b} a^a
       = 0,
   \label{EOM-MSF-cov}
\eea
where $a_a \equiv u_{a;b} u^b$ is the acceleration vector, and $\theta \equiv u^a_{\;\;;a}$ is the expansion scalar; $\theta$ used in this subsection differs from the phase to be introduced later.

In a single component fluid, the fluid quantities defined in Eqs.\ (\ref{Tab-fluid}) and (\ref{fluid-quantities}) are redundant by having three more degrees of freedom. By taking the energy-frame condition $q_a \equiv 0$, without losing generality [the four-vector $u_a$ and the flux vector $q_a$ have three overlapping degrees of freedom], we have $h_a^b \phi_{,b} = 0$, thus
\bea
   \phi_{,a} = - \widetilde {\dot \phi} u_a,
   \label{fluid-MSF-E-frame-u}
\eea
and the fluid quantities and the equation of motion are simplified as
\bea
   & & \mu = {1 \over 2} \widetilde {\dot \phi}{}^2 + V, \quad
       p = {1 \over 2} \widetilde {\dot \phi}{}^2 - V, \quad
       \pi_{ab} = 0,
   \label{fluid-MSF-E-frame} \\
   & & \widetilde {\ddot \phi}
       + \theta \widetilde {\dot \phi}
       + V_{,\phi}
       = 0.
   \label{EOM-MSF-E-frame}
\eea
We note that, the energy-frame condition in Eq.\ (\ref{fluid-MSF-E-frame-u}) is different from the original one of $q_i = 0$ in Eq.\ (\ref{fluid-MSF-cov}), and this difference becomes important for the axion in coherent oscillation stage where we need time average of $q_i$, thus $\langle \widetilde {\dot \phi} h^b_i \phi_{,b} \rangle$, equals to zero. In this way, for example, $\pi_{ab} \neq 0$ in the axion case \cite{Noh-Hwang-Park-2017}; thus, it is safe to use Eqs.\ (\ref{EOM-MSF}) and (\ref{fluid-MSF-cov}), instead. The difference, however, does not appear in the linear order perturbation \cite{Hwang-Noh-2022-oscillation}, and in our case due to slow-motion condition, see below Eq.\ (\ref{EOM-MSF-M}).

%%%%%%%%%%%%%%%%%%%%%%%%%%%%%%%%%%%%%%%%%%%%%%%%%%%%%%%%%%%%%%%
\subsection{Weak-gravity and slow-motion limit}
                                   \label{sec:MSF-WG-SM}

We consider the same approximation used in Sec.\ \ref{sec:WG-SM}; i.e., we consider weak-gravity and slow-motion limits, considering the scalar-type perturbation in zero-shear gauge, and take linear order in perturbed metric.

Using Eq.\ (\ref{four-vector}), fluid quantities in Eqs.\ (\ref{fluid-MSF-E-frame-u}) and (\ref{fluid-MSF-E-frame}) give
\bea
   & & \mu = {1 \over 2 c^2} \dot \phi^2 ( 1 - 2 \alpha )
       + V, \quad
       p = {1 \over 2 c^2} \dot \phi^2 ( 1 - 2 \alpha ) - V,
   \nonumber \\
   & & \phi_{,i} = - {1 \over c^2} \dot \phi v_i, \quad
       \pi_{ij} = 0.
   \label{fluid-MSF-WG}
\eea
The equation of motion in Eq.\ (\ref{EOM-MSF-E-frame}) gives
\bea
   ( 1 - 2 \alpha ) \ddot \phi
       - ( \dot \alpha - 3 \dot \varphi ) \dot \phi
       - c^2 \Delta \phi + c^2 V_{,\phi} = 0,
   \label{EOM-MSF-WG}
\eea
where we used, under the weak-gravity condition,
\bea
   & & g_{00} = - ( 1 + 2 \alpha ), \quad g_{0i} = 0, \quad g_{ij} = ( 1 + 2 \varphi ) \delta_{ij},
   \nonumber \\
   & & g^{00} = - ( 1 - 2 \alpha ), \quad g^{0i} = 0, \quad g^{ij} = ( 1 - 2 \varphi ) \delta^{ij},
   \nonumber \\
   & & \Gamma^0_{00} = {1 \over c} \dot \alpha, \quad \Gamma^i_{j0} = {1 \over c} \dot \varphi \delta^i_j.
\eea
This also follows from the fully nonlinear expression in Eq.\ (117) of \cite{Hwang-Noh-Park-2016}.
We note that Eqs.\ (\ref{fluid-MSF-WG}) and (\ref{EOM-MSF-WG}) are valid to fully nonlinear order in the fluid and field perturbations under the conditions mentioned.

%%%%%%%%%%%%%%%%%%%%%%%%%%%%%%%%%%%%%%%%%%%%%%%%%%%%%%%%%%%%%%%
\section{Axion}
                                   \label{sec:axion}

Now, we consider the axion as a massive scalar field in a coherently oscillating phase. We consider perturbation to nonlinear order. Under the same conditions used in Sec.\ \ref{sec:WG-SM}, the fluid quantities and the equation of motion of the scalar field are derived in Eqs.\ (\ref{fluid-MSF-WG}) and (\ref{EOM-MSF-WG}). For the axion with $V = {1 \over 2} (\omega_c^2/c^2) \phi^2$, with $\omega_c \equiv mc^2/\hbar$ the Compton frequency, we have
\bea
   & & \mu = {1 \over 2 c^2}
       \left[ \dot {\phi}{}^2 ( 1 - 2 \alpha )
       + \omega_c^2 {\phi}{}^2 \right],
   \nonumber \\
   & & p = {1 \over 2 c^2}
       \left[ \dot {\phi}{}^2 ( 1 - 2 \alpha )
       - \omega_c^2 {\phi}{}^2 \right], \quad
       \phi_{,i}
       = - {1 \over c^2} \dot {\phi} v_i,
   \label{fluid-MSF-M} \\
   & & \ddot {\phi} ( 1 - 2 \alpha )
       - ( \dot \alpha - 3 \dot \varphi ) \dot {\phi}
       - c^2 \Delta \phi
       + \omega_c^2 \phi = 0,
   \label{EOM-MSF-M}
\eea
where we may set $\phi ({\bf x}, t) = \bar \phi (t) + \delta \phi ({\bf x}, t)$ with no restriction on the amplitude of $\delta \phi$, and similarly for $A$ and $\theta$ introduced below. As mentioned, in the axion case the energy-frame condition in Eq.\ (\ref{fluid-MSF-E-frame-u}) is not necessarily the same as the original condition of $q_i = 0$, but here due to the slow-motion condition Eq.\ (\ref{fluid-MSF-E-frame-u}) remains valid.

Following \cite{Khmelnitsky-Rubakov-2014}, we take the {\it ansatz}
\bea
   \phi ({\bf x}, t)
       = A ({\bf x}, t)
       \cos{[ \omega_c t + \theta ({\bf x}, t) ]},
   \label{ansatz}
\eea
where we {\it assume} $A$ and $\theta$ are slowly varying in time compared with the Compton frequency, thus strictly {\it ignore} $\dot {A}/ A$ and $\dot {\theta}$ compared with $\omega_c$; reference \cite{Khmelnitsky-Rubakov-2014} assumed $A ({\bf x})$ and $\theta ({\bf x})$ but, as will become clear, the time dependence of $A$ and $\theta$ is necessary for the consistency. Equation (\ref{fluid-MSF-M}) gives
\bea
   & & \mu + ( \bar \mu + \bar p ) \alpha
       = {m^2 c^2 \over 2 \hbar^2} A^2,
   \nonumber \\
   & & p + ( \bar \mu + \bar p ) \alpha
       = - {m^2 c^2 \over 2 \hbar^2} A^2
       \cos{( 2 \omega_c t + 2 \theta )},
   \nonumber \\
   & & v_i = - {\hbar \over m}
       \left( \theta_{,i}
       - { \sin{(2 \omega_c t + 2 \theta)} \over
       1 - \cos{(2 \omega_c t + 2 \theta)} }
       {A_{,i} \over A} \right),
   \label{fluid-M}
\eea
and the sine and cosine parts, respectively, of Eq.\ (\ref{EOM-MSF-M}) give
\bea
   & & 2 {\dot {A} \over A}
       = {\hbar \over m}
       \bigg( \Delta \theta
       + 2 \theta^{,i}
       {A_{,i} \over A} \bigg),
   \label{EOM-M1} \\
   & & \alpha = {\hbar^2 \over 2 m^2 c^2} \bigg(
       {\Delta A \over A}
       - \theta^{,i} \theta_{,i} \bigg)
       + {\hbar \over m c^2} \dot {\theta}.
   \label{EOM-M2}
\eea
Notice that even the equation of motion gives nonlinear relations. For the background pressure we take time average, thus $\bar p = - \bar \mu \langle \cos{(2 \omega_c t + 2 \bar \theta)} \rangle = 0$, and we have $\dot {\bar \theta} = 0$.

Compared with our results up to this point, reference \cite{Khmelnitsky-Rubakov-2014} has {\it ignored} the metric parts, $\alpha$-terms, in Eqs.\ (\ref{ddot-Psi-M}) and (\ref{fluid-M}), and {\it without} addressing the $v_i$ relation in Eq.\ (\ref{fluid-M}) and the equation of motion in Eqs.\ (\ref{EOM-M1}) and (\ref{EOM-M2}), presented
\bea
   & & \mu
       = {m^2 c^2 \over 2 \hbar^2} A^2, \quad
       p
       = - \mu
       \cos{( 2 \omega_c t + 2 \theta )},
   \label{KR1} \\
   & & \Delta \Psi = 4 \pi G \varrho, \quad
       \ddot \Psi = - 4 \pi G \mu
       \cos{( 2 \omega_c t + 2 \theta )}.
   \label{KR2}
\eea
Comparing Eq.\ (\ref{ddot-Psi-M}) and the $p$-relation in Eq.\ (\ref{fluid-M}), apparently, we do not need to ignore the metric to get the $\ddot \Psi$-equation in (\ref{KR2}), and soon we will show that the metric term in the $\mu$-relation of Eq.\ (\ref{fluid-M}) is negligible, but {\it not} in the $p$-relation, see Eq.\ (\ref{alpha-delta}). Including the background fluid quantities in Eq.\ (\ref{KR2}) is not correct in a strict sense as the gravitational potential (metric variable) in Eq.\ (\ref{metric}) is pure perturbed order; it can be justified, though, {\it if} we have the background density negligible compared with the perturbed part as in the galactic dark matter halo or soliton core (composed of the ultra-light axion). Reference \cite{Khmelnitsky-Rubakov-2014} stopped here, without addressing the equation of motion in Eqs.\ (\ref{EOM-M1}) and (\ref{EOM-M2}), the $v_i$-relation in Eq.\ (\ref{fluid-M}), and ignored the fluid representation of dark matter axion.

Equations (\ref{EOM-M1}) and (\ref{EOM-M2}) can be written in more familiar forms. Comparing the ansatz in Eq.\ (\ref{ansatz}) with the following Klein \cite{Klein-1926} and the Madelung \cite{Madelung-1927} transformations
\bea
   \phi = {\hbar \over \sqrt{2m}}
       ( \psi e^{-i \omega_c t} + \psi^* e^{i \omega_c t} ), \quad
       \psi = \sqrt{\varrho \over m} e^{i m u/\hbar},
\eea
respectively, we have
\bea
   & & A = {\hbar \over m} \sqrt{2 \varrho}, \quad
       \theta = - {m \over \hbar} u.
   \label{A-theta}
\eea
Using the new notation, Eqs.\ (\ref{EOM-M1}) and (\ref{EOM-M2}) give Madelung hydrodynamic equations \cite{Madelung-1927}
\bea
   & & \dot \varrho
       + \left( \varrho u^{,i} \right)_{,i}
       = 0,
   \label{E-conserv-axion} \\
   & & \dot u + {1 \over 2} u^{,i} u_{,i}
       + \Phi - {\hbar^2 \over 2 m^2}
       {\Delta \sqrt{\varrho} \over \sqrt{\varrho}}
       = 0,
   \label{Mom-conserv-axion}
\eea
where $\varrho$ and $u$ can be identified as the mass density and velocity potential with ${\bf u} \equiv \nabla u$, respectively.

%%%%%%%%%%%%%%%%%%%%%%%%%%%%%%%%%%%%%%%%%%%%%%%%%%%%%%%%%%%%%%%
\subsection{Constraints from equation of motion}

Let's proceed with these additional relations neglected in \cite{Khmelnitsky-Rubakov-2014}. Our complete set of equations are Eqs.\ (\ref{E-conserv-M})-(\ref{ddot-Psi-M}) for the fluid formulation and Eqs.\ (\ref{fluid-M})-(\ref{EOM-M2}) for the axion counterpart.

Eqs. (\ref{ddot-Psi-M}) and (\ref{fluid-M}) give
\bea
   \ddot \Psi = - 4 \pi G
       \left[ \delta \mu + ( \bar\mu + \bar p ) \alpha \right]
       \cos{( 2 \omega_c t + 2 \theta )}.
   \label{ddot-Psi-alpha}
\eea
By taking time-average for $v_i$-relation in Eq.\ (\ref{fluid-M}), we have $\delta \theta$ determined in terms of the velocity as
\bea
   \theta_{,i} = - m v_i / \hbar,
   \label{theta-v}
\eea
thus, $\delta \theta = m v/\hbar$ with $v_i \equiv - v_{,i}$. The $\mu$-relation in Eq.\ (\ref{fluid-M}) gives
\bea
   {A / \bar A}
       = \pm \sqrt{1 + \delta + \alpha}.
   \label{A-delta}
\eea

Now we consider the equation of motion. The second and third terms in the right-hand side of Eq.\ (\ref{EOM-M2}) vanish in the slow-motion limit [in the third term we used Eq.\ (\ref{Mom-conserv-M}) and the $p$-relation in Eq.\ (\ref{fluid-M}) with time average, and kept only linear order in $\alpha$], and using Eq.\ (\ref{A-delta}) we have
\bea
   \alpha = {\hbar^2 \over 2 m^2 c^2}
       {\Delta A \over A}
       = {\hbar^2 \over 2 m^2 c^2}
       { \Delta \sqrt{1 + \delta + \alpha}
       \over \sqrt{1 + \delta + \alpha} }.
   \label{alpha-A}
\eea
Using Eqs.\ (\ref{theta-v}) and (\ref{A-delta}), Eq.\ (\ref{EOM-M1}) gives
\bea
   \left( \delta + \alpha \right)^{\displaystyle{\cdot}}
       = - \nabla \cdot \left[
       \left( 1 + \delta + \alpha \right) {\bf v} \right].
\eea
Comparison with Eq.\ (\ref{E-conserv-M}), {\it demands}
\bea
   \alpha = - {\delta p / \mu} \ll \delta,
   \label{alpha-delta}
\eea
for consistency; the first relation follows from taking average of $p$-relation in Eq.\ (\ref{fluid-M}). As we have $\alpha \sim {\lambda_c^2 / \lambda^2}$ with $\Delta \sim 1/\lambda^2$ and $c/\omega_c \sim \lambda_c$, this implies
\bea
   {\lambda_c^2 / \lambda^2} \ll \delta.
   \label{condition}
\eea
In the case of linear perturbation, the condition becomes ${\lambda_c^2 / \lambda^2} \ll 1$, \cite{Hwang-Noh-2022-gauges, Hwang-Noh-2022-oscillation}; this condition follows because on sub-Compton scale the axion is not oscillating, thus our ansatz in Eq.\ (\ref{ansatz}) fails to apply. On the galactic halo, we have $\delta \ge 10^4$ \cite{Khmelnitsky-Rubakov-2014}, $\delta$ reaching up to $10^9$ in the soliton core \cite{Schive-etal-2014, deMartino-etal-2017}. Thus, our analysis is valid (consistent) for $\lambda > \lambda_c/\sqrt{\delta} \sim 0.402 {\rm pc}/(m_{22} \sqrt{\delta})$ where $m_{22} \equiv m c^2/(10^{-22} {\rm eV})$. Therefore, from Eqs.\ (\ref{alpha-A}) and (\ref{alpha-delta}), we have
\bea
   \alpha = - {\delta p \over \mu}
       = {\hbar^2 \over 2 m^2 c^2}
       { \Delta \sqrt{1 + \delta} \over \sqrt{1 + \delta} }.
   \label{alpha-p-delta}
\eea

%%%%%%%%%%%%%%%%%%%%%%%%%%%%%%%%%%%%%%%%%%%%%%%%%%%%%%%%%%%%%%%
\subsection{Quantum stress}

Using Eq.\ (\ref{alpha-p-delta}), Eqs.\ (\ref{E-conserv-M})-(\ref{Poisson-eq-M}) give
\bea
   & & \delta \dot \varrho
       + \nabla \cdot
       \left( \varrho {\bf v} \right)
       = 0,
   \label{E-conserv-QS} \\
   & & \dot {\bf v} + {\bf v} \cdot \nabla {\bf v}
       + \nabla \Phi
       = - {\hbar^2 \over 2 m^2}
       \nabla {\Delta \sqrt{1 + \delta} \over \sqrt{1 + \delta}},
   \label{Mom-conserv-QS} \\
   & & \Delta \Phi = 4 \pi G \delta \varrho,
   \label{Poisson-eq-QS}
\eea
where we used the condition in Eq.\ (\ref{condition}). Equations (\ref{E-conserv-QS}) and (\ref{Mom-conserv-QS}) also directly follow from Eqs.\ (\ref{E-conserv-axion}) and (\ref{Mom-conserv-axion}) with ${\bf v} = {\bf u} \equiv \nabla u$, thus the velocity is of the potential type and $v = - u$. The term in the right-hand side of Eq.\ (\ref{Mom-conserv-QS}) is the quantum stress which can be derived from the Schr\"odinger equation as well \cite{Madelung-1927, Hwang-Noh-2022-gauges}.

Equations (\ref{E-conserv-QS})-(\ref{Poisson-eq-QS}) are valid for an axion fluid to fully nonlinear order; the general relativistic counterpart valid to fully nonlinear order in the axion-comoving gauge (in the cosmological context), assuming $H/\omega_c = \hbar H/(m c^2) \ll 1$ (thus, non-relativistic axion assumed), are presented in Eqs.\ (28)-(31) of \cite{Noh-Hwang-Park-2017}; $H \equiv \dot a/a$ is the Hubble-Lema\^itre parameter. The axion contribution remains the same and the only relativistic contributions appear in the metric nonlinearities.

To the linear order, using Eq.\ (\ref{alpha-p-delta}), Eq.\ (\ref{ddot-delta-eq}) gives
\bea
   \ddot \delta - 4 \pi G \bar \varrho \delta
%       = \Delta {\delta p \over \varrho}
%       = - c^2 \Delta \alpha
       = - {\hbar^2 \Delta^2 \over 4 m^2} \delta,
   \label{ddot-delta-eq-M}
\eea
thus the Jeans criterion remains the same as in expanding background \cite{Khlopov-1985}. The competition between gravity and quantum stress terms gives the quantum Jeans scale
\bea
   \lambda_J = {2 \pi \over k_J}
%       = {\pi c \over \sqrt{ \omega_c \sqrt{\pi G \varrho}}}
       = \pi \sqrt{\hbar \over m \sqrt{\pi G \bar \varrho}},
   \label{Jeans-scale}
\eea
where $k$ is the wavenumber with $\Delta \equiv - k^2$. Post-Newtonian approximation, including the leading $H/\omega_c$ order relativistic correction for the axion, is studied in \cite{Hwang-Noh-2023-PN}.

%%%%%%%%%%%%%%%%%%%%%%%%%%%%%%%%%%%%%%%%%%%%%%%%%%%%%%%%%%%%%%%
\subsection{Quantum oscillation}

Using Eqs.\ (\ref{theta-v}) and (\ref{alpha-delta}), Eq.\ (\ref{ddot-Psi-alpha}) gives
\bea
   {\ddot \Psi \over c^2} = - 4 \pi G \delta \varrho ({\bf x}, t)
       \cos{\left[ 2 \omega_c t + 2 \bar \theta
       + 2 {m \over \hbar} v({\bf x}, t) \right]}.
   \label{ddot-Psi-eq}
\eea
This is the result presented in \cite{Khmelnitsky-Rubakov-2014} with the perturbed phase now determined in terms of the velocity potential as $2 \bar \theta + 2 m v({\bf x}, t)/\hbar$; the background phase $\bar \theta$ is a constant in space and time. The $v$-term includes the position dependent phase and the post-Newtonian order frequency shift.

By {\it ignoring} the time dependence of $\delta \varrho$ and $v$, we can derive a solution \cite{Hwang-Noh-2022-oscillation}
\bea
  & &\Psi = - {G \delta \varrho \over \pi} \lambda^2
       + {G \delta \varrho \over 4 \pi} \lambda_c^2
       \cos{\left( 2 \omega_c t + 2 \bar \theta
       + 2 {m \over \hbar} v \right)}
   \nonumber \\
   & & \quad
       = - {G \delta \varrho \over \pi} \lambda^2
       + {G \delta \varrho \over 4 \pi} \lambda_c^2
       \Big[ \cos{\left( 2 {m \over \hbar} v \right)}
       \cos{( 2 \omega_c t + 2 \bar \theta )}
   \nonumber \\
   & & \quad \qquad
       - \sin{\left( 2 {m \over \hbar} v \right)}
       \sin{( 2 \omega_c t + 2 \bar \theta )} \Big],
   \label{ddot-Psi-sol}
\eea
where $\lambda_c \equiv h/(mc)$ is the Compton wavelength and the non-oscillatory solution is the one supported by density inhomogeneity in Eq.\ (\ref{Poisson-eq-M}). We may set \cite{Khmelnitsky-Rubakov-2014}
\bea
   & & \Psi ({\bf x}, t)
       \equiv \Psi_\delta ({\bf x})
       + \Psi_c ({\bf x}) \cos{(2 \omega_c t + 2 \bar \theta)}
   \nonumber \\
   & & \qquad
       + \Psi_s ({\bf x}) \sin{(2 \omega_c t + 2 \bar \theta)},
   \label{Psi-solution} \\
   & & \Psi_\delta
       = - 4 \left( {\lambda \over \lambda_c} \right)^2
       \overline \Psi,
       \quad
       \overline \Psi
       \equiv {G \delta \varrho \over 4 \pi} \lambda_c^2,
   \nonumber \\
   & &
       \Psi_c = \overline \Psi
       \cos{\left( 2 {m \over \hbar} v \right)}, \quad
       \Psi_s = - \overline \Psi
       \sin{\left( 2 {m \over \hbar} v \right)}.
       \label{PsiPsi}
\eea
The oscillation frequency and dimensionless amplitude of the oscillating parts are
\bea
   & & f = 2 \nu_c = 4.84 \times 10^{-8} m_{22} {\rm Hz},
   \nonumber \\
   & & {\overline \Psi \over c^2}
       = {G \bar \varrho \lambda_c^2 \over 4 \pi c^2} \delta
       = 2.09 \times 10^{-23} \Omega_{0.25} h_{0.7}^2 m_{22}^{-2} \delta,
\eea
where $H \equiv 70 h_{0.7} {\rm km/sec/Mpc}$, $\Omega = 8 \pi G \bar \varrho/(3 H^2)$ and $\Omega_{0.25} \equiv \Omega/(0.25)$.

On the other hand, using $v_i = - v_{,i} = u_{,i}$, Eq.\ (\ref{Mom-conserv-axion}) gives
\bea
   \dot v = {1 \over 2} v^2
       + \Phi - {\hbar^2 \over 2 m^2}
       {\Delta \sqrt{\varrho} \over \sqrt{\varrho}},
\eea
where $v^2 \equiv v^i v_i$. {\it Assuming} a stationary medium, thus ignoring time dependence of the right-hand side, we have
\bea
   & & v ({\bf x}, t) = \left( {1 \over 2} v_0^2
       + \Phi - {\hbar^2 \over 2 m^2}
       {\Delta \sqrt{\varrho} \over \sqrt{\varrho}} \right) t
       + v_0 ({\bf x})
   \nonumber \\
   & & \qquad
       \equiv \delta v ({\bf x}) t + v_0 ({\bf x}).
   \label{delta-v}
\eea
Using $2 \omega_c t + 2 \bar \theta + 2 m v/\hbar = 2 \omega_c ( 1 + \delta v/c^2 ) t + 2 \bar \theta + 2 m v_0/\hbar$, we have the frequency shift
\bea
   {\delta \omega \over \omega_c}
       \equiv {1 \over c^2} \delta v
       = {1 \over c^2} \left( {1 \over 2} v_0^2
       + \Phi - {\hbar^2 \over 2 m^2}
       {\Delta \sqrt{\varrho} \over \sqrt{\varrho}} \right).
       \label{delta_omega}
\eea
In this way, part of the perturbed phase causes frequency shift. The three dimensionless terms are the velocity dispersion, gravitational potential, and quantum stress affecting the frequency shift. Apparently, these are of the first post-Newtonian orders, and are quite small \cite{Hwang-Noh-2023-PN}.

%%%%%%%%%%%%%%%%%%%%%%%%%%%%%%%%%%%%%%%%%%%%%%%%%%%%%%%%%%%%%%%
\section{Implications for PTA}
                                  \label{sec:PTA}

In Ref.\cite{Khmelnitsky-Rubakov-2014}, it was shown that the presence of an oscillating gravitational potential affects the travel time of radio beams coming from pulsars.
Following the same procedure, the delay produced by the oscillating gravitational potential in Eq.\ (\ref{ddot-Psi-sol}) on the Times of Arrival (TOAs) of the radio pulses emitted by the pulsar $p$ and registered by Pulsar Timing Array (PTA) experiments is:
\bea
   & & \hskip -.4cm
       \Delta t(t)
       = - \frac{\overline \Psi}{\omega_c} \sin\bigg\{ \omega_c \bigg[
       {\delta \omega_e - \delta \omega_p
       \over \omega_c} t +
       \left( 1 + {\delta \omega_p \over \omega_c}
       \right) \frac{D}{c}
   \nonumber \\
   & & \quad \hskip -.4cm
       + \frac{1}{c^2} \left[ v_0 ({\bf x}_e)
       - v_0 ({\bf x}_p) \right] \bigg] \bigg\}
   \nonumber \\
   & & \quad \hskip -.4cm
       \times
       \cos\bigg\{ \omega_c \bigg[ 2 \left( 1
       + {\delta \omega_e + \delta \omega_p
       \over 2 \omega_c} \right) t
       - \left( 1 + {\delta \omega_p \over \omega_c} \right) \frac{D}{c}
   \nonumber \\
   & & \quad \hskip -.4cm
       + \frac{1}{c^2} \left[ v_0 ({\bf x}_e) + v_0 ({\bf x}_p) \right] \bigg]
       + 2 \bar \theta \bigg\},
	\label{delta-t}
\eea
where ${\bf x}_e$ is the Earth position, ${\bf x}_p$ the pulsar position and $D$ is the Earth-pulsar distance. $\delta \omega_e/\omega_c$ and  $\delta \omega_p/\omega_c$ are the velocity potential-induced frequency shifts at the Earth and at the pulsar, respectively [see Eq. (\ref{delta_omega})]; we ignored $\delta \omega/\omega_c$ correction in the amplitude. In the derivation, we assumed a constant dark matter density, \textit{i.e.} $\overline{\Psi} ({\bf x}_p) = \overline{\Psi} ({\bf x}_e) = \overline{\Psi}$ and we expanded the velocity potential consistently with Eqs.\ (\ref{delta-v}) and (\ref{delta_omega}).
Noticeably, both the phase and the amplitude appearing in Eq.\ (\ref{delta-t}) depend on the velocity potential.
In particular, the ultra-light axion model predicts that the PTA signature must be correlated with the dark-matter velocity field.

For a typical velocity {300\,{\rm km/s}} at the length scale $\lambda\simeq {\rm kpc}$, we estimate
\bea
   2 {\omega_c v \over c^2} = 2 {m v \over \hbar}
       \sim 4 \pi {\nabla \cdot {\bf v} \over c \lambda_c \Delta}
       \sim 5.1 m_{22} \lambda_{\rm kpc} v_{300},
   \label{v-parameter}
\eea
where $\lambda_{\rm kpc} \equiv \lambda/(1000 {\rm pc})$ and $v_{300} \equiv |{\bf v}|/(300 {\rm km}/{\rm sec})$. Adding this contribution on top of the recently observed signal from PTAs \cite{Antoniadis_2023rey, NANOGrav_2023gor, Parkes_2023} might open up an interesting opportunity to provide sensible constraints not only on the dark matter density, as recently done in \cite{Smarra_2023, Antoniadis_2023, Afzal_2023}, but also on the common phase $\bar\theta$. However, it must be noticed that the current uncertainties on pulsar distance measurements [$\sim \mathcal{O}(0.1\div1\text{kpc})$] \cite{Verbiest_2012} might make it challenging to disentangle this formulation from the previous one in Ref.\cite{Khmelnitsky-Rubakov-2014}, as they will effectively give rise to a pulsar-dependent random phase. Therefore, current PTA experiments may not allow for discriminating between the two scenarios, but future facilities (e.g. the Square Kilometer Array) will provide much clearer insights \cite{Smits:2011zh}.
% We encourage future work to shed light on these aspects.
Moreover, if we have no information about the velocity potential $v$ along the line of sight of pulsars and the background common global phase $\bar \theta$, we may regard these as random unknown parameters.
In that case, by introducing
\bea
   \alpha ({\bf x}) \equiv {\omega_c \over c^2} v_0 ({\bf x}_e)
       + \bar \theta, \quad
       \alpha ({\bf x}_p)
       \equiv {\omega_c \over c^2} v_0 ({\bf x}_p)
       +  \bar \theta,
\eea
Eq.\ (\ref{delta-t}) is written exactly as Eq.\ (3.4) in Ref.\cite{Khmelnitsky-Rubakov-2014} (neglecting the post-Newtonian order frequency shifts) and the analysis would follow what previously done  \cite{Porayko-2014, deMartino-etal-2017, Porayko-2018, Smarra_2023}.
In this respect, the {\it Gaia} mission, with precise astrometric measurement of the angular positions, proper motions in the sky, parallaxes, and radial velocities of nearby stars \cite{GAIA} may have potential to probe the velocity structure of the dark matter halos relevant to cover the pulsars used in PTA surveys and thus could allow us to distinguish the two scenarios. For recent attempt in that direction, see \cite{Lim-2023}.

%%%%%%%%%%%%%%%%%%%%%%%%%%%%%%%%%%%%%%%%%%%%%%%%%%%%%%%%%%%%%%%
\section{Discussion}
                                     \label{sec:discussion}

We present a rigorous derivation of the oscillating gravitational potential caused by oscillating axion pressure perturbation in Eq.\ (\ref{ddot-Psi-eq}). Compared with previous work in \cite{Khmelnitsky-Rubakov-2014}, we determined the phase in terms of the perturbed velocity potential. The velocity potential has both the frequency shift and the phase specific to individual pulsar. The frequency shift is the post-Newtonian effect and includes the velocity dispersion, gravitational potential, and the quantum stress, see Eq. (\ref{delta_omega}). The TOAs registered
by PTA experiments is presented in Eq.\ (\ref{delta-t}). Our results are valid in the fully nonlinear matter and field inhomogeneities under the approximation of weak-gravity and slow-motion. We considered scalar-type perturbation in zero-shear gauge.

Currently, the PTA measurements of NANOGrav \cite{NANOGrav_2023gor}, EPTA \cite{Antoniadis_2023rey}, or PPTA \cite{Parkes_2023} clearly hint at a first detection of a stochastic nano-hertz frequency gravitational wave background by recovering an angular pattern consistent with the Hellings-Downs curve \cite{Hellings:1983fr}. The most promising origin of such a nano-hertz background lies in SuperMassive Black-Hole Binaries (SMBHBs), although the observed signal is in mild tension with some n{\"a}ive astrophysical assumption (e.g. circularity of the binary orbit) \cite{NANOGrav:2023hfp,Phinney:2001di, Antoniadis_2023}.

However, a plethora of alternative explanations are also possible \cite{NANOGrav:2023hvm,EPTA:2023xxk,EPTA:2023xiy,Cyr:2023pgw}. Although the current dataset shows no strong evidence for anisotropies \cite{NANOGrav:2023tcn} nor gravitational waves from individual SMBH binaries \cite{NANOGrav:2023pdq}, combining observations in IPTA and accumulating more data would allow an in-depth study of nano-hertz band gravitational waves.

Furthermore, accurate distance measurements of the pulsars monitored by PTAs may enable us to extract meaningful information from the so-called pulsar term, whose phase is commonly assumed as a random parameter. In such a case, cross-correlating the PTA results with the distribution of known astrophysical quantities, such as dark-matter density distribution and dark-matter velocity distribution, will play an important role in confirming the gravitational-wave origin of the observed time offset from PTAs. Our result in this paper will be essential for such a test, in particular, against the possibility that the time offset comes from the oscillating light scalar fields.

%%%%%%%%%%%%%%%%%%%%%%%%%%%%%%%%%%%%%%%%%%%%%%%%%%%%%%%%%%%%%%%
\section*{Acknowledgments}

C.S.\ acknowledges support from the European Union's H2020 ERC Consolidator Grant ``GRavity from Astrophysical to Microscopic Scales" (Grant No. GRAMS-815673) and the EU Horizon 2020 Research and Innovation Programme under the Marie Sklodowska-Curie Grant Agreement No. 101007855. C.S.\ also acknowledges hospitality from the Galileo Galilei Institute (GGI), where part of his contribution was carried on.
D.J.\ was supported by KIAS Individual Grant PG088301 at Korea Institute for Advanced Study. 
H.N.\ was supported by the National Research Foundation (NRF) of Korea funded by the Korean Government (No.\ 2021R1F1A1045515).
J.H.\ was supported by IBS under the project code IBS-R018-D1 and by the NRF of Korea (No.\ NRF-2019R1A2C1003031).

%%%%%%%%%%%%%%%%%%%%%%%%%%%%%%%%%%%%%%%%%%%%%%%%%%%%%%%%%%%%%%%%%
%
%   References
%
%%%%%%%%%%%%%%%%%%%%%%%%%%%%%%%%%%%%%%%%%%%%%%%%%%%%%%%%%%%%%%%%%

%%%%%%%%%%%%%%%%%%%%%%%%%%%%%%%%%%%%%%%%%%%%%%%%%%%%%%%%%%%%%%%

%%%%%%%%%%%%%%%%%%%%%%%%%%%%%%%%%%%%%%%%%%%%%%%%%%%%%%%%%%%%%%%
\end{document}